\documentclass[fdp,a4paper,fleqn%
]{w-art}
\usepackage{times,cite,w-thm}
\usepackage{amsmath}
\usepackage{mathptmx}
\usepackage{multirow}
\newcommand{\eq}[1]{\begin{equation}
                     \begin{split} #1 \end{split}
                     \end{equation}}

\newcommand{\ov}{\overline}
\newcommand{\p}{\partial}
\newcommand{\tri}{\hspace{1.5pt}\triangle\hspace{1.5pt}}

\def\BNT{\,\hbox{\hbox to -0.2pt{\vrule height 6.5pt width .2pt\hss}\rm N}}
\def\BRT{\,\hbox{\hbox to -0.2pt{\vrule height 6.5pt width .2pt\hss}\rm R}}
\def\BZ{{\rm Z{\hbox to 3pt{\hss\rm Z}}}}
\def\IP{\relax{\rm I\kern-.18em P}}
\def\BCT{\,\hbox{\hbox to -3pt{\vrule height 6.5pt width .2pt\hss}\rm C}}
\def\BCS{\,\hbox{\hbox to -2.2pt{\vrule height 4.5pt width .2pt\hss}$
    \scriptstyle\rm C$}}
\def\BCSS{\,\hbox{\hbox to -2pt{\vrule height 3.3pt width
    .2pt\hss}$\scriptscriptstyle \rm C$}}
\def\BC{{\mathchoice{\BCT}{\BCT}{\BCS}{\BCSS}}}
\theoremstyle{plain}

\theoremstyle{definition}

\usepackage[]{graphicx}
\begin{document}
\DOIsuffix{theDOIsuffix}
\pagespan{1}{}



\hfill MPP-2014-54

\title[Noncommutative Geometry in String Theory]{A Course on Noncommutative Geometry in String Theory}


\author[Ralph Blumenhagen]{Ralph Blumenhagen\inst{1,}%
\footnote{Lecture notes for a 3h course given at the Workshop on
Noncommutative Field Theory and Gravity September 8 - 15, 2013, Corfu}
}
\address[\inst{1}]{Max-Planck-Institut f\"ur Physik, F\"ohringer Ring 6, 80805 M\"unchen, Germany}
\begin{abstract}
In this pedagogical mini course
the basics of the derivation of
the noncommutative structures appearing in string theory
are reviewed. First we discuss  the well established
appearance of the noncommutative Moyal-Weyl star-product 
in the correlation functions of open string vertex operators 
on a magnetized D-brane. Second, we will review the most recent
attempts to generalize these concepts to the closed string
moving in a nongeometric flux background. 
\end{abstract}

\maketitle                   






\section{Introduction}\label{ra_sec1}

It is well known that Einstein's theory of gravity, i.e. general
relativity, cannot be consistently quantized with the rules
developed in the framework of quantum field theory. Using
perturbation theory around a flat metric, the ultraviolet
divergences require infinitely many counter-terms.

Thus, it is natural to consider theories in which a natural
minimal length is imposed. With the lessons
from quantum theory, where a minimal volume element in phase space   
is introduced, one could imagine introducing noncommuting
coordinates to generate a minimal volume element in space-time.
This is the idea of noncommutative geometry, which by now 
is mathematically well developed and whose physical applications
is the subject of this workshop. What is still lacking is
a concrete idea how to use this formalism to formulate
a consistent theory of quantum gravity. In fact noncommutative geometry
has mainly been employed for applications to generalized
gauge theories, like the noncommutative standard model of 
Connes-Lott \cite{Connes:1990qp}.

Another approach to introduce a minimal length is by postulating
that the fundamental physical objects are extended so that their
interaction also violates the principle of locality. Certainly, 
the most prominent example is string theory, for which the 
existence of gravity is a direct consequence and which contains
infinitely many new ultraviolet degrees of freedom, namely
all the higher oscillation modes of the string, making the 
string theory on-shell Feynman diagrams finite.

The question is whether the above mentioned two approaches
are related or maybe even complementary to each other.
Since string theory is by far better developed, one can 
approach this question by looking for the appearance
of noncommutative target-space structures in both open and
closed string theory.  For the open string there exists a well established
relation, namely open strings ending on a D-brane supporting
a non-vanishing two-form (gauge flux) see a noncommutative
D-brane world-volume geometry. In fact, as shown by 
Seiberg-Witten \cite{Seiberg:1999vs}
one can describe in a certain limit the effective theory on the brane
as a noncommutative gauge theory. Then, the star-product in
Witten's open string field theory \cite{Witten:1985cc} is directly
related to the Moyal-Weyl star-product for functions on the target space.

During the recent years there have been some attempts to reveal  
similar noncommutative structures in closed string theory 
\cite{Blumenhagen:2010hj,Lust:2010iy,Blumenhagen:2011ph,Condeescu:2012sp,
  Mylonas:2012pg, Chatzistavrakidis:2012qj, Andriot:2012vb,Bakas:2013jwa,Mylonas:2013jha,Blumenhagen:2013zpa}.
As will be explained in this course, in this case the analysis
is far more subtle and seems to be obscured by the  limitation
of  only having available  a background dependent formulation of string theory.
That means that the  background satisfies the string equations of motion, which
by definition are the constraints for conformal invariance of
the sigma model action. There are certain indications that
the closed string is related to the appearance of not only noncommutative but
even  nonassociative target space geometries. It is fair to say that the 
precise relation is still not completely settled, but a couple
of interesting observation have been reported.

In this mini-course a pedestrian introduction into the appearance
of noncommutaitve geometry in both the open and the closed
string sector is provided. It is aimed not for specialists in string
theory but rather for people interested in noncommutative geometry.
This  is why the course  starts at a considerable  basic level.
Moreover, a partial overlap with the already existing 
proceeding article \cite{Blumenhagen:2011yv}
could not be avoided.

\section{String theory preliminaries}

In this section, let us introduce some basic material of the bosonic
closed and open string. We will restrict ourselves to those aspect
which will become relevant in the course of this lecture. For more
details and derivations we refer the reader to the existing text books
\cite{Polchinski:1998rq, Polchinski:1998rr, Blumenhagen:2013fgp}.

\vspace{0.5cm}
\noindent
{\bf Aspects of closed string world-sheet theory}
\vspace{0.2cm}

The usual approach to string theory is in a first quantized version, i.e.
one considers a string moving in a target space background with metric 
$g_{ab}$, Kalb-Ramond field $B_{ab}$ and dilaton $\Phi$, whose
dynamics is governed by a two-dimensional non-linear sigma model.
With $\Sigma$ denoting the world-sheet of the closed string, its action reads
\begin{equation}
  \label{action_730178}
\mathcal S_{\rm bulk}=-\frac{1}{4\pi\alpha'}\int_{\Sigma} d\sigma d\tau\, \left(h^{ij}\, g_{ab}(X)\, \p_i X^a\p_j X^b
+\epsilon^{ij}\, B_{ab}(X)\, \p_{i}X^a\p_{j}X^b\right)\, ,
\end{equation}
where $h^{ij}$ denotes the Minkowski world-sheet metric and
we suppressed  the dilaton part. 
Here the coordinate fields depend on the word-sheet coordinates
$(\sigma,\tau)$ and satisfy the boundary condition 
$X^a(\sigma+2\pi,\tau)=X^a(\sigma,\tau)$ for closed strings.
Choosing for instance 
the simple background of a flat metric $g_{ab}=\eta_{ab}$ and $B_{ab}=0$,
the 2D equation for motion is simply the wave equation
$(\partial^2_\tau-\partial^2_\sigma) X^a=0$. Introducing  light-cone coordinates
$u=\tau+\sigma$, $v=\tau-\sigma$, this becomes $\partial_u \partial_v X^a=0$.
Its solution splits into left and right moving waves $X^a(\sigma,\tau)=
X^a_L(u)+X^a_R(v)$ with the mode expansion
\eq{
\label{modeexpand}
           X^a_L(u)&={x_0^a\over 2}+{\alpha'\over 2} P^a (\tau+\sigma) + i
\sqrt{\alpha'\over 2}\sum_{n\ne 0} {\alpha_n\over n} e^{-in (\tau+\sigma)}\, ,\\
X^a_R(v)&={x_0^a\over 2}+{\alpha'\over 2} P^a (\tau-\sigma) + i\sqrt{\alpha'\over 2}
\sum_{n\ne 0} {\overline\alpha_n\over n} e^{-in (\tau-\sigma)}
}
where $x_0^a$ denotes the center of mass position and $P^a$
the center of mass momentum operator of the closed string. One also defines
$\alpha^a_0=\overline\alpha^a_0=\sqrt{\alpha'\over 2} P^a$.
The modes satisfy the  commutation relations
\eq{
  [\alpha_m^a,\alpha^b_n]=\eta^{ab}\, m\, \delta_{m,-n}\,, \qquad
  [\overline\alpha_m^a,\overline\alpha^b_n]=\eta^{ab}\, m\, \delta_{m,-n}
\,, \qquad
  [\alpha_m^a,\overline\alpha^b_n]=0
\, .
}
Thus, one can define the modes $\alpha^a_n$ and $\overline\alpha^a_n$ with $n<0$ 
as raising operators and the positive modes as lowering operators.
The Hilbert space 
is then given by the Fock-space of the two algebras acting on a ground state
$\vert \vec p\rangle$ with $P^a\vert \vec p\rangle=p^a \vert \vec p\rangle$
subject to the level matching constraint $N_L-N_R =0$ with $N_{L/R}$ denoting
the left and right moving occupation numbers. The target space
mass of such a state is given by ${\alpha' \over 2} M^2=N_L+N_R -2$.

For utilizing the powerful methods of complex analysis, one Wick rotates
the world-sheet into Euclidean signature and defines $z=\exp(\tau+i\sigma)$.
From the commutator algebra one can compute the two-point function
as
\eq{
     \langle X^a(z,\overline z)\, X^b(w,\overline w)\rangle=-\alpha'\, \eta^{ab}
     \log\vert z-w\vert \, .
}
For each state in the Hilbert space $\vert\phi\rangle$, there exist 
a corresponding 
field $\phi(z,\overline z)$, where the correspondence is determined via
$\lim_{z\to 0} \phi(z)\vert 0\rangle=\vert\phi\rangle$.
For instance, the ground state tachyon of mass ${\alpha' \over 2}M^2=-2$ and momentum
$\vec p$ corresponds to the field (vertex operator)
$V_T=\exp(i\vec p\cdot \vec X(z,\overline z))$
with ${\alpha' \over 2}\vec p^2=2$. At the first  excited level one finds the on-shell
vertex operator
\eq{  V_G= \xi_{ab} \partial X^a(z) \,\overline\partial X^b(\overline z)\, 
    \exp(i\vec p\cdot \vec X(z,\overline z))
}
with $\vec p^2=0$ and transverse polarization, i.e. $p^a \xi_{ab}=p^b
\xi_{ab}=0$. For symmetric polarization this is the graviton/dilaton and
for anti-symmetric one the Kalb-Ramond field. 
The tree level closed string scattering amplitude for $N$ such on-shell vertex
operators is defined as
\eq{
     {\cal A}\sim \int d^2 z_1\ldots d^2 z_N\   \prod_{i=1}^3 
        \delta(z_i-z_i^0)\ |z_{12} z_{13} z_{23}|^2 \ 
       \langle V_1(z_1,\overline z_1)\dots  V_1(z_N,\overline z_N)\rangle_{S^2}
}
where we included the universal contribution from the ghosts and
 already implemented the effect of the $SL(2,\BC)$ invariance, which allows
to move the first  three points to three fixed ones $z_i^0$.   
The most common choice is $z_1=\infty, z_2=1, z_3=0$.  
The correlation function below the integral is to be computed
in the conformal field theory on the Riemann sphere (compactified
complex plane). Higher loop amplitudes involve CFT correlation
functions on higher genus Riemann surfaces.

Note that in the mode expansion  \eqref{modeexpand} the zero mode part is
left-right symmetric. This changes when one considers the compactication
of the string on e.g. a circle of radius $R$ in the $a=25$ direction. 
In this case, starting with the general ansatz
$X^{25}_{L/R}(u)={x_0^a\over 2}+{\alpha'\over2} p_{L/R}^{25} (\tau\pm\sigma) +\ldots$,
one has quantized Kaluza-Klein momentum $(p^{25}_L+p^{25}_R)/2=m/R$, 
$m\in \BZ$ and
quantized winding strings $X^{25}(\tau,2\pi)=X^{25}(\tau,0)+2\pi n R$, $n\in \BZ$.
Thus, one gets 
\eq{
        p^{25}_{L/R}={m\over R} \pm {nR\over \alpha'}\, \quad{\rm with\ 
mass\ spectrum}\quad 
   \alpha' M^2=\alpha' {m^2\over R^2} +{1\over \alpha'} n^2 R^2 +\ldots\, .
}
This is  invariant under the transformation
\eq{
    T:\; R\to {\alpha'\over R}\, , \qquad m\leftrightarrow n\, .
}
This acts on the left and right momenta as $T:(p^{25}_L,p^{25}_R)\to (p^{25}_L,-p^{25}_R)$
and can be extended to the entire string theory. Thus, closed
strings on a circle admit a new type of symmetry, so called T-duality, which
acts on the coordinates in a left-right asymmetric way 
$T:(X^{25}_L,X^{25}_R)\to (X^{25}_L,-X^{25}_R)$.  

\vspace{0.5cm}
\noindent
{\bf Target space equations of motion}
\vspace{0.2cm}

From the computation of string scattering amplitudes one can deduce
an effective target space action for the massless modes, i.e. the graviton, the
dilaton and the Kalb-Ramond field. 
Alternatively, one can start from the non-linear sigma model 
and require that on-shell backgrounds are given by its  conformal
fixed points. From the 2D point of view, the coupling constants
are given by the target space background fields. 
This quantum field theory is treated perturbatively in a dimensionless
coupling $\sqrt{\alpha'}/R$, where $R$ is a characteristic length scale
of the background. 
At leading order, the beta-function equations for the couplings
$g_{\mu\nu}$, $B_{\mu\nu}$ and $\Phi$ read
\begin{eqnarray}
  \label{targeteom}
 0 &=&\beta_{ab}^G  =  \alpha' \Bigl( R_{ab}-\frac{1}{4}\: H_{a}{}^{cd}\, H_{bcd}  
  +2 \nabla_a\nabla_b\Phi\Bigr)
    +O({\alpha'}^2)\nonumber\, ,\\
 0 &=&\beta_{ab}^B  =  \alpha'\Bigl( -{1\over 2}  \nabla_c H^c{}_{ab} +
   \alpha' H_{ab}{}^c \nabla_c \Phi \Bigr)
    +O({\alpha'}^2)\, ,\\
   0 &=&\beta_{ab}^\Phi  =  {1\over 4} (d-d_{\rm crit})+ \alpha' \Bigl(
    (\nabla \Phi)^2 -{1\over 2} \nabla^2 \Phi -{1\over 24} H^2 \Bigl)
    +O({\alpha'}^2)\, .\nonumber
\end{eqnarray}
The first equation, at leading order in $\alpha'$, is nothing else than 
Einstein's equation with sources.
Clearly, in this approach one is assuming from the very beginning that
the string is moving through a Riemannian geometry with additional smooth
fields. However, it is well known that there exist conformal
field theories which cannot be identified with such simple
geometries. These are  left-right asymmetric like for instance
asymmetric orbifolds. The latter are asymmetric at some
orbifold fixed points but one can imagine asymmetric CFTs which
are not even locally geometric. We will see that 
the target space interpretation of such asymmetric CFTs is closely
related to noncommutative geometry.

\vspace{0.5cm}
\noindent
{\bf Aspects of open strings}
\vspace{0.2cm}

Now consider open strings with $0\le \sigma\le \pi$. Such open strings
can end on D-branes carrying in general a non-vanishing gauge flux
$F_{ab}=\partial_a A_b - \partial_b A_a$ on their world-volume. 
One extends  the  non-linear 
world-sheet
sigma-model action \eqref{action_730178} to include the gauge fields on the 
D-branes at the end of the open string world-sheet
\eq{
\label{nonlinearffield}
\mathcal \mathcal S=\mathcal S_{\rm bulk}
-\int_{\p \Sigma} d\tau\ A_a(X)\, \p_\tau X^a \ .}
The action has the Abelian gauge invariance of the vector potential at the
boundary $\delta A_a = \p_a \Lambda$
and the combined two-form gauge invariance of the antisymmetric
tensor $B_{ab}$, which also involves a boundary term,
\eq{\label{deltaBA}
\delta B_{ab} = \p_a \zeta_b - \p_b \zeta_a, \, \qquad
\delta A_a = -\frac{1}{2\pi\alpha'} \zeta_a \ .}
For constant metric $G_{ab}$, antisymmetric tensor $B_{ab}$ and gauge field
strength $F_{ab}$ the action \eqref{nonlinearffield} simplifies.
First we notice that, using
\eq{\label{actionB}
\int_{\Sigma} d^2\sigma\,  \epsilon^{ij}\, B_{ab}\, 
\p_{i}X^a\p_{j}X^b = \int_{\p \Sigma}d\tau\, B_{ab}\, X^a\p_{\tau}X^b\, ,}
the term involving $B_{ab}$ can be written as a boundary term.
For constant $F_{ab}$ one can also write
\eq{\label{actionA}    
\int_{\p \Sigma}d\tau\, A_{a}\, \p_{\tau}X^a 
={1\over 2} \int_{\p \Sigma}d\tau\, F_{ab}\, X^a\, \p_{\tau}X^b}
so that the total world-sheet action becomes
\begin{equation}
\label{nonlinearffieldb}
\mathcal S=-\frac{1}{4\pi\alpha'}\int_{\Sigma} d\sigma d\tau\, \eta^{ij}\, g_{ab}\, \p_i X^a\p_j X^b 
-\frac{1}{4\pi\alpha'}\int_{\p \Sigma}d\tau \bigl(B_{ab}+(2\pi \alpha')\, F_{ab}\bigr) \, X^a\p_{\tau}X^b\; .
\end{equation}
Therefore, the end points of the open string couple
to the gauge invariant field strength 
\eq{\label{calF}
2\pi\alpha' {\cal F}_{ab} = B_{ab} + 2\pi\alpha' F_{ab}\ .}
The boundary conditions at $\sigma=0,\pi$ that follow from the variation of
the world-sheet action are 
\eq{\label{mixedbc}
g_{ab}\, \p_\sigma X^b + (2\pi\alpha' {\cal F}_{ab})\ \p_\tau X^b \vert_{\sigma=0,\pi}= 0 \ .}
For the target space $T^2$ with flat metric 
$g_{ab}=\delta_{ab}$ and the constant field strength 
${\cal F}={\cal F}_{12}$, we obtain the so-called mixed boundary conditions 
\eq{\label{mixedbctorus}
\p_\sigma X + (2\pi\alpha' {\cal F})\,  \p_\tau Y &= 0 \;, \\
\p_\sigma Y - (2\pi\alpha' {\cal F})\,  \p_\tau X &= 0 \; .
}
Thus, the gauge field interpolates between a Neumann boundary condition
$\p_\sigma X^a=0$ and a Dirichlet boundary condition $\p_\tau X^a$.

\begin{figure}
\centering
\includegraphics[width=0.25\textwidth]{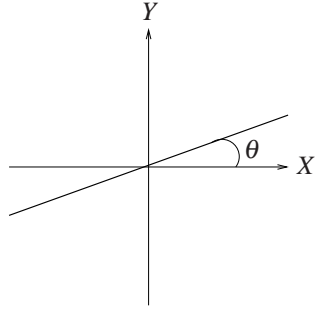}
\begin{picture}(0,0)
\put(-20,56){$\theta$}
\put(0,50){$X$}
\put(-58,108){$Y$}
\end{picture}
\vskip5pt
\caption{D-brane at angle}\label{fig_intersect}
\end{figure}

For an open string with either pure Neumann or pure Dirichlet boundary conditions
on both ends, one obtains the following mode expansion
\eq{
\label{modeexpandopen}
           X^a(z)_{\rm NN/DD}&={x_0^a}+2\alpha' P^a\,\log |z| + i
\sqrt{2\alpha'} \sum_{n\ne 0} {\alpha_n\over n} (z^{-n}\pm \overline z^{-n})
\quad {\rm with} \ \,
[\alpha_m^a,\alpha^b_n]=\eta^{ab}\, m\, \delta_{m,-n}\,. \
}
Thus, relative to the closed string the degrees or freedom are halved.
The open string two-point function for these two cases can be straightforwardly computed
as
\eq{
\label{twopointopen}
     \langle X^a(z)\, X^b(z')\rangle_{{\rm NN/DD}}=-\alpha' \delta^{ab}\left(
     \log\vert z-z'\vert\pm \log\vert z-\overline z'\vert \right)\, 
}
where $z$ and $z'$ take values in the complex upper  half plane. 
Now we discuss how T-duality acts in the open string sector.
Since it exchanges momentum and winding,  a T-duality in the $X$-direction
acts as $T: \p_\tau X\leftrightarrow \p_\sigma X$.  
Let us consider a  D-brane on non-compact $\,\BC$, 
which is shown in figure \ref{fig_intersect}.
The D-brane intersecting the $X$-axis under an angle $\theta$ 
provides the following boundary condition for an open string
\eq{
\label{boundaryintersecting}
\cos\theta\,\p_{\sigma}X+\sin\theta\,\p_{\sigma}Y&=0 \, ,\\
-\sin\theta\,\p_{\tau}X+\cos\theta\,\p_{\tau}Y&=0 \; .}
Applying a T-duality in the $Y$-direction, these boundary 
conditions become
\eq{
\label{boundarymagnetised}
\p_{\sigma} X+ \tan\theta\,\p_{\tau}Y=0\, , \\
\p_{\sigma}Y -\tan\theta\, \p_{\tau} X= 0\,.
} 
These are precisely of the mixed type \eqref{mixedbctorus} so that
we can identify $(2\pi\alpha' {\cal F})=\tan\theta)$.
Under T-duality in the $Y$- direction a $D1$-brane at angle
is mapped to a magnetized $D2$-brane. 
Moreover, a (left-right symmetric) rotation of the $D1$-brane by an
angle $\theta$
\eq{ 
\vec X_L\to A\, \vec X_L, \quad\quad \vec X_R\to  A\,\vec X_R \, ,
\qquad {\rm with}\qquad
A=\left(\begin{matrix} \cos\theta & \sin\theta \\
             -\sin\theta & \cos\theta \end{matrix}\right)
}
is mapped under T-duality to a left-right asymmetric rotation defined as
\eq{ 
\label{asymmrot}
\vec X_L\to A\, \vec X_L, \quad\quad \vec X_R\to  A^T\, \vec X_R \, .
}
Thus we summarize that left-right asymmetric symmetries of the world-sheet
act in the open string sector via changing the magnetic flux
on the brane.

\section{Noncommutative geometry: open strings}
\label{sec_open}

\begin{figure}
\centering
\includegraphics[width=0.40\textwidth]{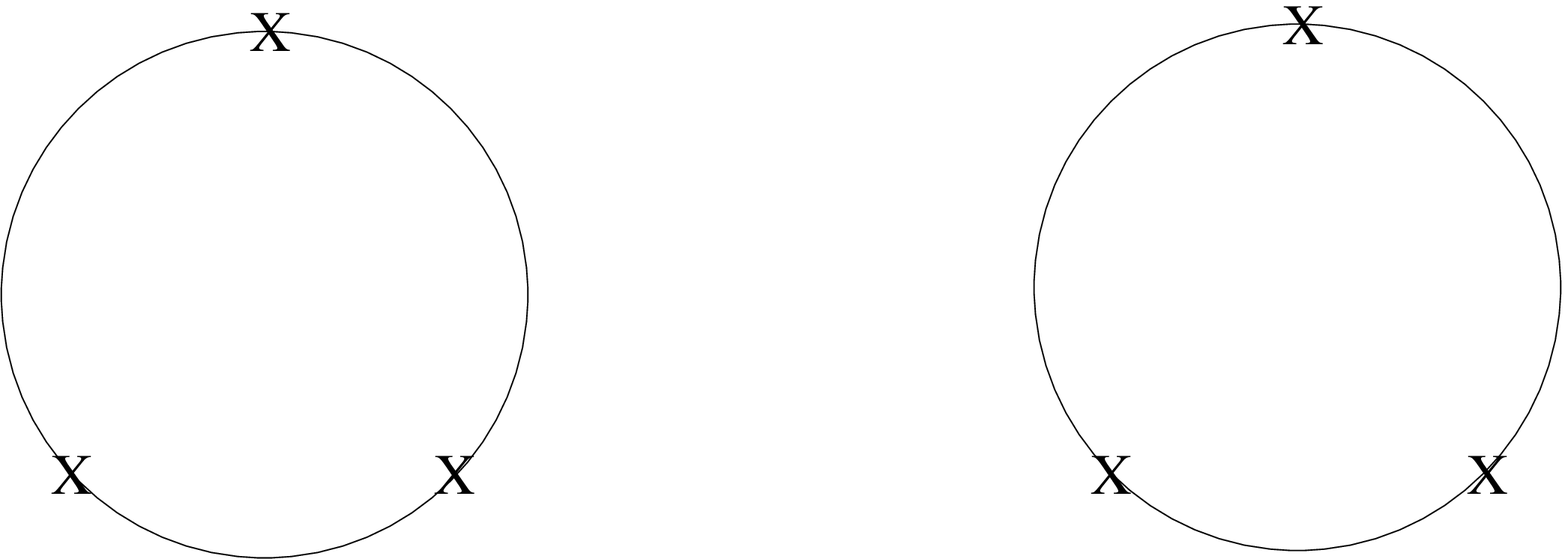}
\begin{picture}(0,0)
\put(-146,60){$\infty$}
\put(-35,60){$\infty$}
\put(-197,8){$X^a(\tau)$}
\put(-118,8){$X^b(\tau')$}
\put(-83,8){$X^b(\tau)$}
\put(-7,8){$X^a(\tau')$}
\put(-108,50){${\cal F}_{ab}\ne 0$}
\put(-90,30){$\ne$}
\end{picture}
\vskip5pt
\caption{Disc diagram}\label{fig_discs}
\end{figure}

That noncommutative geometry is related to open string theory has long
been suspected. For instance, in Witten's cubic open string field \cite{Witten:1985cc}
the fundamental product between two string fields is noncommutative,
though still associative and cyclic. Expanding the theory around
the flat background with vanishing Kalb-Ramond field, this noncommutativity
is not visible in the zero mode sector, i.e. the target space coordinates
are still commutative. However, by turning on a background which is
sensitive to the order of two end-points of an open string, one
has a chance to also make manifest the noncommutativity in the zero mode
sector. As we will derive in this section, such a background
is given by a constant ${\cal F}_{ab}$ flux.
Mathematically this means that for non-vanishing  ${\cal F}_{ab}$ 
the two disc correlators shown in figure \ref{fig_discs} are different.
In the following we will compute this amplitude using the methods
presented in the previous section and which were introduced in \cite{Blumenhagen:2000fp}.

\vspace{1cm}
\noindent
We are still working on a two-dimensional flat target space and
recall the two-point function \eqref{twopointopen} 
on the upper half plane  for NN boundary conditions 
\eq{
\langle X^a(z)\, X^b(z') \rangle&=
                 -\alpha'\delta^{ab}\left( \log |z-z'| +\log |z-\ov{z}'| \right)
                 \\
                &=-\alpha'\delta^{ab}{1\over 2}
                \left( \log (z-z') + \log (\ov{z}-\ov{z}')
             +\log (z-\ov{z}') + \log (\ov{z}-{z}')\right) \, .
}
In the second line we have split the two-point function so that, 
formally using 
\eq{ X^a(z)=X^a_L(z) + X^a_R(\ov{z}), 
}
we can directly read off the individual contributions from
the left- and right-movers. According to our previous discussion, we can
turn on a non-vanishing 
constant flux by performing an asymmetric rotation
\eqref{asymmrot}.
A straightforward computation leads to the 
following expression
for the propagator in the rotated coordinates
\eq{  \langle X^a(z)\, X^b(z') \rangle=
                 -\alpha'\delta^{ab} \log |z-z'| - \alpha'\delta^{ab}
               &\left( \sin^2\theta-\cos^2\theta\right) \log |z-\ov{z}'| \\
               &\qquad -\alpha' \epsilon^{ab} \sin\theta\, \cos\theta\,
               \log\left({z-\ov{z}'\over \ov{z}-z' } \right) .
}
Using $\tan\theta=B_{12}=B$ we can write this as
\eq{  
\label{hannover96}
\langle X^a(z)\, X^b(z') \rangle=
                 -\alpha'\delta^{ab} \Big(\log |z-z'| -\log |z-\overline z'|\Big)
&- {2\alpha'\over 1+B^2}\delta^{ab}\, \log |z-\ov{z}'| \\
               &\qquad -{\alpha' B\over 1+B^2} \epsilon^{ab} 
\log\left({z-\ov{z}'\over \ov{z}-z' } \right) .
}
Finally, restricting this to the real axis, $z=\overline z=t$ and
$z'=\overline z'=t'$, and choosing the branch cut of the
logarithm in the second line of \eqref{hannover96} 
along the negatively imaginary axis one obtains
\eq{ 
\label{twopointboundary}
 \langle X^a(t)\, X^b(t') \rangle=
- {\alpha'\over 1+B^2}\delta^{ab}\, \log (t-t')^2 
            +{i\pi \alpha' B\over 1+B^2} \epsilon^{ab} \epsilon(t-t')\, .
}
Here $\epsilon(x)=1$ for $x$ positive and $\epsilon(x)=-1$ for $x$ 
negative. Note that the second term is not symmetric under exchange
of $X^a$ and $X^b$ and therefore is the source of noncommutativity.
This expression agrees precisely  with the propagator
  derived in  \cite{Seiberg:1999vs} with the identification of the open string
metric 
\eq{
G^{ab}={1\over 1+B^2}\delta^{ab} \qquad {\rm  and}\qquad
\theta^{ab}=2\pi\alpha'{B\over 1+B^2}\epsilon^{ab}\, .
}
Thus, by applying an asymmetric rotation we have found an elegant and short
way of deriving this  propagator.
This relation of the open and closed string quantities can be compactly written
as
\eq{
\label{openclosed}
     \left(G^{-1}+{1\over 2\pi\alpha'}\theta\right)=\left(g+2\pi\alpha'{\cal F}\right)^{-1}\, .
}

Using the
correlator \eqref{twopointboundary}, one can compute 
the operator product expansion of open string tachyon vertex operators 
at the  boundary 
\eq{ e^{i p X}(t)\,   e^{i q X}(t') =
           (t-t')^{2\alpha' G^{ab}\, p_a\, q_b }\,
            {\rm exp}\left(-{i\over 2} \theta^{ab}\, p_a\, q_b\right) \ 
             e^{i (p+q) X}(t')+\ldots\quad {\rm for}\ \ t>t'\, .
}
The extra phase is independent of the word-sheet coordinates and
can therefore be considered as a target-space effect. Indeed, this
phase can be considered to arise due to a noncommutative deformation
of the product of functions on the target space, namely from the  Moyal-Weyl
star-product
\eq{
\label{starproduct}
   (f\,\star\, g)(x)= \exp\biggl(
   {i\over 2}\, \theta^{ab}\,
      \partial^{x_1}_{a}\,\partial^{x_2}_{b} \biggr)\, f(x_1)\, g(x_2)\Bigr|_{x} \;.
}
This noncommutative product satisfies a couple of properties:
\begin{enumerate}
\item{For the coordinate functions $f=x^a$ and $g=x^b$ it leads to
the star-product commutator $[x^a,x^b]_\star=i\,\theta^{ab}$, the defining
relation of the Moyal-Weyl plane.}
\item{The star-product is associative $f\star (g\star h)=(f\star g)\star h$.
This is consistent with the associativity of the operator product expansion
in conformal field theory. Actually, in CFT one requires  crossing
symmetry of correlation functions, which implies the weaker
constraint of associativity up to boundary term, i.e. 
$\int dx f\star (g\star h)-\int dx (f\star g)\star h=0$.}
\item{It is also cyclic in the sense $\int dx\, f\star g=\int dx\, g\star f$
and similarly for cyclic permutations of the product of $N$ functions.
This is a consequence of the fact that  
the conformal $SL(2,\BRT)$ symmetry group leaves
the cyclic order of the inserted vertex operators invariant.}
\end{enumerate}

So far we considered the most tractable case of a constant background
flux in flat space, but mathematically 
the star-product was defined in \cite{Kontsevich:1997vb} 
for an  arbitrary  Poisson structure $\theta^{ij}$. In this case
the corresponding  star-product is still associative.
However  one can also consider 
the same product for only a quasi Poisson structure, which 
then leads to a nonassociative star-product (see e.g. \cite{Mylonas:2012pg}).
Physically, also in this most generic situation, as long as
the background satisfies the string equations of motion, i.e.
for an on-shell background, the properties 1.-3. above should still
be satisfied. Following essentially \cite{Cornalba:2001sm, Herbst:2001ai,Blumenhagen:2013zpa},
let us confirm this for the first non-trivial order terms.

Now we have the  physical situation of an open string ending on a D-brane
with generic non-constant $B$-field, i.e. non-vanishing
field strength $H$ supported on a brane embedded in some
generically curved space.
At leading order in derivatives this leads to 
a noncommutative product
\eq{
     f\circ g=f\cdot g + 
  i\, {1\over 2} \theta^{ab}\, \partial_a f\, \partial_b g -
   &{1\over 8} \theta^{ab}\theta^{cd}\, \partial_a\partial_c f\, \partial_b\partial_d g  \\ &
-{1\over 12} \big(\theta^{am}\partial_m \theta^{bc}\big) \big(
  \partial_a\partial_b f \, \partial_c g + \partial_a\partial_b g\, \partial_c f\big) \ldots\, 
}
where the first line is just the expansion of the Moyal-Weyl product
\eqref{starproduct}.
The associator for this product becomes 
\eq{
\label{openasso}
    (f\circ g)\circ h-f\circ (g\circ h)={1\over 6}\,
      \theta^{abc}\, \partial_a f \,\partial_b g \,\partial_c h+O(\partial\theta^2)
}
with $\theta^{abc}=3\,\theta^{[\underline{a}m}\partial_m
  \theta^{\underline{bc}]}$, which by definition vanishes for a Poisson tensor.
This can be written as 
\eq{
    \theta^{abc}=\theta^{am} \theta^{bn}
      \theta^{cp} 
   (2\pi\alpha' dF+H)_{mnp}=
 \theta^{am} \theta^{bn}
      \theta^{cp} H_{mnp}
}
where 
we used  the Bianchi identity $d{F}=0$.

For a slowly varying gauge field, the effective action is 
given by the Dirac-Born-Infeld (DBI)  action
\eq{
    S_{\rm DBI}=\int dx  \sqrt{g+2\pi\alpha' {\cal F}}\, .
}
Varying it with respect to the gauge potential $A$ in $\,2\pi\alpha'{\cal
  F}=B+2\pi\alpha' dA$,
one gets the equation of motion
\eq{ 
\label{eoma}
 \partial_a\left( \sqrt{g+2\pi\alpha'{\cal F}} \left[ (g+2\pi\alpha' {\cal
     F})^{-1}\right]^{[\underline{ab}]}\right) ={1\over 2\pi\alpha'}\partial_a\left( \sqrt{g+2\pi\alpha'{\cal F}} \, \theta^{ab}\right)=0
}
where we have used \eqref{openclosed}.
Then, it directly follows that up to leading  order in $\partial \theta$
the $\circ$-product satisfies  cyclicity
\eq{
\label{intiprop}
     \int dx \sqrt{g+2\pi\alpha' {\cal F}}\, f\circ g= \int dx
     \sqrt{g+2\pi\alpha' {\cal F}}\, g\circ f
\, .}
Indeed, e.g. at order $O(\theta)$   the difference between the left and the right hand side is
a total derivative on-shell
\eq{
     {i\over 2}   \int dx \sqrt{g+2\pi\alpha'{\cal F}}\, \theta^{ab}\, \partial_a f\, \partial_b g
    = {i\over 2} \int dx\, \partial_a \!\left(\sqrt{g+2\pi\alpha' {\cal F}}\, \theta^{ab}\, f\, \partial_b g\right) =0\, 
}  
Thus, as expected from CFT, the 
product of two functions is cyclic, once
the background satisfies the string equations of motion.

Similarly, the associator below the integral also gives a total
derivative at leading order in $\partial \theta$
\eq{
\label{assoprop}
        \int dx \sqrt{g+2\pi\alpha' {\cal F}}\; \Big( &(f\circ g)\circ h-f\circ
        (g\circ h)\Big)\\
&={1\over 6}
   \int dx \sqrt{g+2\pi\alpha' {\cal F}}\; \Big( 
 \theta^{am} \theta^{bn}
      \theta^{cp} H_{mnp} \, \partial_a f \,\partial_b g \,\partial_c h+O(\partial\theta^2)\Big)\\
&={1\over 6}
\int dx\, \partial_a \!\left(\sqrt{g+2\pi\alpha' {\cal F}}\, \theta^{abc}\,
f\, \partial_b g\, \partial_c h\right) +O(\partial\theta^2)\, .
} 
Thus, we conclude that, as expected from the open string conformal
field theory, on-shell (up to $O(\partial\theta^2)$) 
the $\circ$-product is associative up to boundary
terms.
Our argument seems to suggest that it
might be relevant for string field theory in off-shell backgrounds,
but to our knowledge this has not been made explicit.

\section{Noncommutative geometry: closed string generalization}

In the previous section we have seen that noncommutativity arises 
for open strings in a magnetic flux background
leading to noncommutative gauge theories. 
Clearly, if noncommutative geometry has any significance for improving the
ultraviolet behavior of quantum gravity, it should also appear
for closed strings.
Thinking about this question, one realizes that 
the closed string analogue must  be different as
here two vertex operators are inserted in the bulk of 
a two-sphere $S^2$ and no unambiguous ordering of $z_1$ and $z_2$ 
can be defined.
Therefore, one does not expect the same kind of noncommutativity to arise.

\begin{figure}[ht]
\centering
\includegraphics[width=0.26\textwidth]{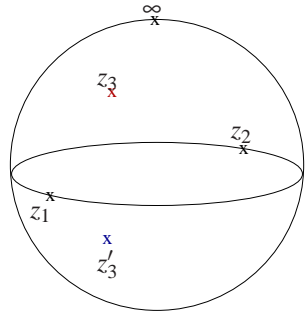}
\begin{picture}(0,0)
\put(-63,111){$\infty$}
\put(-105,35){$z_1$}
\put(-30,65){$z_2$}
\put(-80,85){$z_3$}
\put(-80,15){$z'_3$}
\end{picture}
\vskip5pt
\caption{Geometric four punctured sphere}\label{fig_sphere1}
\end{figure}

However, as  shown in figure \ref{fig_sphere1}, adding one more point might improve the situation.
Just looking at the $S^2$ geometrically, $z_1$ and $z_2$ define 
a geodesic. Then adding $z_3$ one can distinguish the two situation
that it lies on the same or opposite hemisphere as the point $\infty$.
Then following the open string logic, turning on a three-vector
flux $\Theta^{abc}$ distinguishing these two configurations can make
this tri-product structure manifest.
Thus, in this case one would e.g. expect a fundamental non-vanishing
tri-product of the form\footnote{One could also try to not consider 
this tri-product to be fundamental but the result of a non-vanishing
Jacobiator for a generalized star-product. This approach was 
initiated in \cite{Lust:2010iy} and is discussed in the 
lecture by P. Schupp (see the lecture notes \cite{Mylonas:2014aga} and also \cite{Mylonas:2012pg,Andriot:2012vb,Bakas:2013jwa,Mylonas:2013jha}).} 
\begin{equation}
\label{threebracketcon}
   (f_1 \tri\, f_2\, \tri\, f_3)(x) = \exp\Bigl(
   \Theta^{abc}\,
      \partial^{x_1}_{a}\,\partial^{x_2}_{b}\,\partial^{x_3}_{c} \Bigr)\, f_1(x_1)\, f_2(x_2)\,
   f_3(x_3)\Bigr|_{x} \;
\end{equation}
which allows to define a totally antisymmetric tri-bracket 
\begin{equation}
\label{antisymtripcon}
   \bigl [f_i,f_j,f_k \bigr]=\sum_{\sigma\in S_3} {\rm sign}(\sigma) \;  
     f_{\sigma(i)}\, \tri\,  f_{\sigma(j)}\, \tri\,  f_{\sigma(k)} \; ,
\end{equation}
where $S_3$ denotes the permutation group of three elements.

\begin{figure}[ht]
\centering
\includegraphics[width=0.26\textwidth]{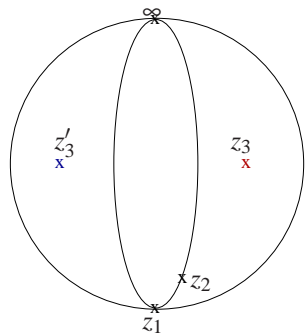}
\begin{picture}(0,0)
\put(-63,111){$\infty$}
\put(-63,-6){$z_1$}
\put(-45,10){$z_2$}
\put(-30,62){$z_3$}
\put(-96,62){$z'_3$}
\end{picture}
\vskip5pt
\caption{CFT  sphere diagram}\label{fig_sphere2}
\end{figure}

However, there is a caveat to this simple picture, namely that
in string theory one has to take into account 
the conformal $SL(2,\BC)$ symmetry of the world-sheet,.
As shown in figure \ref{fig_sphere2}, this allows
to map the three points $z_1$, $z_2$ and the north pole
to $z_1=0$, $z_2=1$ and $z_4=\infty$, i.e. they do lie on the
same geodesic. Therefore, adding $z_3$ does not lead to any
clearly distinguished order. This is consistent with
the fact that in CFT a four-point function can be expressed
in terms of the $SL(2,\BC)$ invariant cross-ratio and
features crossing symmetry. The latter is nothing else than
associativity of the operator product expansion.

\noindent
From these simple arguments we arrive at the two features
for the generalization of noncommutativity to the closed
string sector:
\begin{enumerate}
\item{Noncommutativity/nonassociativity
 is expected to rather involve a three-vector $\Theta^{abc}$
and a corresponding tri-product. }
\item{However, its effect is expected to be invisible in the CFT.}
\end{enumerate}

\noindent
Clearly, with these insights one could stop at this stage. However, since
these ideas have the potential to  be a window into off-shell string theory
or at least into structures present in theories beyond the usual
approach to string theory,
let us proceed and investigate how precisely the two features 1. and 2. 
above are realized in closed string theory.

First we will clarify what the nature of the three-vector flux
$\Theta^{abc}$ might be. In fact in the study of nongeometric
fluxes, precisely such a candidate appeared, the so-called
$R$-flux. Since the appropriate framework for describing it
is double field theory we will also present some of its structure.
For a more detailed presentation we refer the reader
to the review articles \cite{Aldazabal:2013sca, Berman:2013eva, Hohm:2013bwa}.

Second, we will perform a CFT analysis for a constant
flux background and, in analogy to the open string story,
will compute correlation functions of vertex operators.

\vspace{0.5cm}
\noindent
{\bf The nature of the flux $\Theta^{abc}$}
\vspace{0.2cm}

In closed string theory there generically exists a massless Kalb-Ramond
two form, whose field strength $H=dB$ is a three-form. At first sight,
this seems to be the natural candidate for $\Theta^{abc}$. 
Whereas a constant two-form flux on a D-brane only slightly
changes the boundary conditions of the open string, due to
\eqref{targeteom}, a constant
$H$-flux really backreacts on the geometry and also 
implies that the underlying exact CFT is not any longer free but becomes
highly interacting. Therefore, this situation is much harder to analyze.

However, the $H$-flux is not the only candidate for a source of
nonassociativity. Applying
T-duality  to the  closed string background \cite{Shelton:2005cf} given by
a flat space with constant non-vanishing 
three-form flux $H=dB$, results in a background with geometric 
flux. This so-called  twisted torus is still a conventional string background,
but a second T-duality leads to a  nongeometric flux background.
These are spaces in which  the
transition functions between two charts of a manifold are allowed
to be T-duality transformations, hence they are also 
called T-folds \cite{Hellerman:2002ax,Dabholkar:2002sy}.
After formally applying a third T-duality, not along an isometry direction 
anymore,
one obtains an $R$-flux background which does not 
admit a clear target-space interpretation. It was proposed 
that this background does not correspond to an ordinary geometry even locally,
but instead gives rise to a nonassociative 
geometry \cite{Bouwknegt:2004ap}. Therefore, this nongeometric flux
seems to be the prime candidate for $\Theta^{abc}$.

As mentioned, for the last T-duality giving $R$-flux one cannot
apply the Buscher rules. During the last years a formalism was developed
which allows to precisely perform such T-dualities. This is called
double field theory (DFT) and is a field theory for the
massless modes of the bosonic string which has manifest
$O(D,D)$ symmetry. A full introduction into this fast developing
subject is beyond the scope of this lecture and we refer to
\cite{Aldazabal:2013sca,Berman:2013eva, Hohm:2013bwa}
for more details.
Here we just give a very few ingredients which allow to perform
the series of T-dualities mentioned above.

The main idea is that one doubles the coordinates $X^A=(x^a,\tilde x_a)$
by introducing so-called winding coordinates $\tilde x_a$.
This allows to put the diffeomorphism symmetry and the $B$-field
gauge transformations on equal footing by combining
the dynamical fields $g_{ab}$ and $B_{ab}$  
in a  generalized metric
\eq{
\label{genmetric}
    {\cal H}_{AB}=
      \left(\begin{matrix}  g^{ab} &  -g^{am}B_{mb} \\
            B_{am}g^{mb} & g_{ab} -B_{am}g^{mn}B_{nb} \end{matrix}\right) \, .
}
The  global $h\in O(D,D)$ symmetry acts as
\eq{
\label{dft_trafo}
         {\cal H}'&=h^t\, {\cal H} h \,, \qquad\qquad
         X'=h\, X\,, \qquad\quad \partial'=(h^t)^{-1} \,\partial \;. 
}
One can construct an $O(D,D)$ invariant action for this generalized
metric, where eventually one halves the degrees of freedom
by imposing the so-called strong constraint
\eq{
  \label{strong_c}
   \partial_a\partial^a=0\,, \qquad
    \partial_a f\, \tilde \partial^a g +\tilde\partial_a f\, \partial^a g
    =0\,  
}
where $f,g$ denote the fundamental objects in DFT like
${\cal H}_{AB}$ and gauge parameters.

In order to see how T-duality is described in this framework consider
a flat torus $T^3$ with a constant $H$-flux with $B_{12}=h
x^3$.
The generalized metric then takes the form\footnote{Except for the last step, the $R$-flux, the following
computation can be found in the appendix of \cite{Andriot:2011uh}.}
\eq{
    {\cal H}=\left(\begin{array}{ccc|ccc}
1 & 0 & 0 & 0 & -h x^3 & 0\\
0 & 1 & 0 & h x^3 &  0 & 0\\
0 & 0 & 1 & 0 & 0 & 0\\\hline
0 & h x^3 & 0 & 1+(hx^3)^2 & 0 & 0\\
-h x^3 & 0 & 0 & 0 & 1+(hx^3)^2  & 0  \\
0 & 0 & 0  & 0 & 0 & 1
\end{array}\right)\, .
}
A T-duality in the $x_1$ direction is a special $O(D,D)$ transformation
and acts via conjugation
${\cal H}'={\cal T}^t_1 {\cal H} {\cal T}_1$ with the matrix
\eq{
    {\cal T}_1=\left(\begin{array}{ccc|ccc}
    0 & 0 & 0 &  1 & 0 & 0 \\
    0 & 1  & 0  & 0 & 0 & 0 \\
    0 & 0 & 1  & 0 & 0 & 0 \\\hline
    1 & 0 & 0 & 0 & 0 & 0 \\
    0 &  0 & 0 & 0 & 1 & 0 \\
    0 & 0 & 0 & 0 & 0 & 1
\end{array}\right)\, .
}
One gets
\eq{
    {\cal H}'=\left(\begin{array}{ccc|ccc}
    1+(h x^3)^2 & h x^3 & 0 &  0 & 0 & 0 \\
    h x^3  & 1  & 0  & 0 & 0 & 0 \\
    0 & 0 & 1  & 0 & 0 & 0 \\\hline
    0 & 0 & 0 & 1 & -hx^3 & 0 \\
    0 &  0 & 0 & -h x^3 & 1+(h x^3)^2 & 0 \\
    0 & 0 & 0 & 0 & 0 & 1
\end{array}\right)
}
which corresponds to the metric
\eq{
    ds^2= (dx^1-h x^3 dx^2)^2 +(dx^2)^2 + (dx^3)^2
}
on the twisted three torus with
vanishing $B$-field. Thus, this background is characterized by a
constant geometric flux $F^1{}_{23}=h$.

Second, let us perform a T-duality
in the isometric $x_2$ direction. The 
generalized metric transforms as
${\cal H}''={\cal T}^t_2 {\cal H}' {\cal T}_2$ with
\eq{
    {\cal H}''=\left(\begin{array}{ccc|ccc}
 1+(h x^3)^2 & 0 & 0 & 0 & h x^3 & 0\\
0 &  1+(h x^3)^2 & 0 & -h  x^3 &  0 & 0\\
0 & 0 & 1 & 0 & 0 & 0\\\hline
0 & -h  x^3 & 0 & 1 & 0 & 0\\
h  x^3 & 0 & 0 & 0 & 1  & 0  \\
0 & 0 & 0  & 0 & 0 & 1
\end{array}\right)\, .
}
Reading off the metric and the B-field, we find
\eq{
    g=\left(\begin{array}{ccc}
 {1\over 1+(h x^3)^2} & 0 & 0 \\
0 &  {1\over 1+(h x^3)^2} & 0 \\
0 & 0 & 1 
\end{array}\right)\, ,\qquad
  B=\left(\begin{array}{ccc}
 0 & -{h x^3\over 1+(h x^3)^2} & 0  \\
 {h x^3 \over 1+(h x^3)^2} & 0 & 0 \\
0 & 0 & 0
\end{array}\right)\, .
}
Due to the terms in the denominators, this looks a bit awkward.
However, in generalized geometry and DFT one can also parametrize 
the generalized metric  via
\eq{
\label{nongeomframe}
    {\cal H}_{ab}=
      \left(\begin{matrix} \tilde g^{ab}-\beta^{am} \tilde g_{mn}\beta^{nb} &
        \beta^{am}  \tilde g_{mb}\\
                  - \tilde g_{am} \beta^{mb} & \tilde g_{ab}\end{matrix}\right)
}
where $\beta={1\over 2}\beta^{ab}\,\partial_a\wedge \partial_b$ denotes
an anti-symmetric bi-vector.
The geometric frame \eqref{genmetric} and this nongeometric one
are  related via the field redefinition
\eq{
            \tilde g&=g-Bg^{-1}B\, ,\\
            \beta &=- \tilde g^{-1} B g^{-1}\, ,
}
which is reminiscent of the Buscher rules and the open string 
relation \eqref{openclosed}. 
Using this frame one gets the simple result
\eq{
    \tilde g=\left(\begin{array}{ccc}
 {1} & 0 & 0 \\
0 &  {1} & 0 \\
0 & 0 & 1 
\end{array}\right)\, ,\qquad
  \beta=\left(\begin{array}{ccc}
 0 & h x^3 & 0  \\
 {-h x^3} & 0 & 0 \\
0 & 0 & 0
\end{array}\right)\, .
}
Then one defines the nongeometric $Q$ flux as
\eq{
    Q_3{}^{12}=\partial_3 \beta^{12}=h
}
so that this background has a flat metric $\tilde g$ and a constant $Q$-flux.

Thus, we have established a chain of  T-duality transformations 
\eq{
  H_{ijk} \;\leftrightarrow\;
   F_{ij}{}^{k} \;\leftrightarrow\;
  Q_{i}{}^{jk} \;\leftrightarrow\;
  R^{ijk} \; 
}
where we have indicated the final step, namely the result of
performing a T-duality also in the non-isometric $x_3$ direction.
We note that, according to the $O(D,D)$ action  \eqref{dft_trafo} for
${\cal T}_3$,
the normal and winding coordinate are exchanged as 
$x_3\leftrightarrow \tilde x_3$ so that
\eq{
    \tilde g=\left(\begin{array}{ccc}
 {1} & 0 & 0 \\
0 &  {1} & 0 \\
0 & 0 & 1 
\end{array}\right)\, ,\qquad
  \beta=\left(\begin{array}{ccc}
 0 & h \tilde x_3 & 0  \\
 {-h \tilde x_3} & 0 & 0 \\
0 & 0 & 0
\end{array}\right)\, .
}
Then one defines the nongeometric $R$ flux as
\eq{
    R^{123}=3\, \tilde\partial^{[3} \beta^{12]}=h\, .
}
Therefore, this flux explicitly involves winding derivatives and
is not visible in ordinary geometry.

\vspace{0.5cm}
\noindent
{\bf CFT analysis of $\Theta^{abc}$ flux}
\vspace{0.2cm}

Now, we will analyze the conformal field theory for a closed
string moving on a flat space with  a constant $H$-flux background.
The aim is to perform a similar computation as for the
open string, i.e. in particular  to compute
$N$-point functions of tachyon vertex operators.
But before we can do that, we need to define a CFT.
The essential observation is that a Ricci flat metric, vanishing dilaton  and a
constant $H$-flux solves the string equation of motion up to linear
order in $H$. 
Since we are interested in a linear order effect, it is appropriate to
perform conformal perturbation theory up to  linear order in $H$ 
around the trivial background. 
Therefore, the starting point is a  flat  metric and a constant
$H$-flux  specified by 
\begin{equation}
  \label{setup_01}
   ds^2=\sum_{a=1}^N \bigl(dX^a\bigr)^2, \hspace{20pt} 
   H=  \frac{2}{{\alpha'}^2}\,  \Theta_{abc}\,  dX^a\wedge dX^b\wedge dX^c\; ,
\end{equation}
where for simplicity  we focus  on  $N=3$.
As said, the expectation is that this background  
corresponds to a CFT up to linear order in $H$. 
Moreover, since the origin of T-duality lies in conformal field 
theory where it is nothing
else than an asymmetric reflection $(X_L,X_R)\to (X_L,-X_R)$,
one can  try to analyze the $R$-flux case also
from the CFT point of view.
The remainder of this article  is essentially a brief version of the 
more exhaustive analysis originally presented in  \cite{Blumenhagen:2011ph}
and reviewed in \cite{Blumenhagen:2011yv}.

To proceed, we write  the action \eqref{action_730178} as the sum of a 
free part $\mathcal S_0$ and a perturbation $\mathcal S_1$. 
Choosing a  gauge   such that $B_{ab} = \frac{1}{3}  H_{abc}\, X^c$, we have
\begin{equation}
  \label{action_pertub_01}
  \mathcal S = \mathcal S_0 + \mathcal S_1 \hspace{20pt}{\rm with}\hspace{20pt}
  \mathcal S_1=\frac{1}{2\pi\alpha'} \:\frac{H_{abc}}{3} \int_\Sigma d^2 z\, X^a
  \partial X^b \,\overline\partial X^c \;.
\end{equation}
We expect  ${\cal S}_1$ to be a marginal operator (only) up to linear order
in $H$. A correlation function of $N$ operators $\mathcal O_i[X]$ can be computed via the path integral in the usual way 
\eq{
  \label{pi_01}
  \bigl\langle \mathcal O_1 \ldots \mathcal O_N \bigr\rangle
  = \frac{1}{\mathcal Z} \int [dX]
  \,
    \mathcal O_1 \ldots \mathcal O_N \,e^{-\mathcal S[X]} 
  \,,
}
where $\mathcal Z$ denotes the vacuum functional given by $\mathcal Z =\int
[dX] \,e^{- S[X]}$. In the limit of small fluxes, we can do conformal
perturbation theory and expand \eqref{pi_01}  
in the perturbation $\mathcal S_1$ leading to
\eq{
  \label{master_01}
  \bigl\langle  \mathcal O_1 \ldots \mathcal O_N \bigr\rangle 
  =\quad&\bigl\langle  \mathcal O_1 \ldots \mathcal O_N\bigr\rangle_{0} 
  - \bigl\langle  \mathcal O_1 \ldots \mathcal O_N\: \mathcal S_1 \bigr\rangle_{0} \\[2mm]
  +\:& \frac{1}{2}\, \Bigl[\,
  \bigl\langle  \mathcal O_1 \ldots \mathcal O_N\: \mathcal S^2_1 \bigr\rangle_{0} 
  - \bigl\langle  \mathcal O_1 \ldots \mathcal O_N \bigr\rangle_{0} \times
  \bigl\langle  \mathcal S^2_1 \bigr\rangle_{0} \,\Bigr]
   +    \mathcal O\bigl( H^3\bigr) \;.
}
First, we compute the correction
to the three-point functions of three currents $J^a=i\partial X^a$, 
$\overline J^a=i\partial \overline X^a$. It turns out that 
there are also non-vanishing correlators like 
$\langle J^a J^b \overline J^c\rangle$, i.e. the currents are not 
holomorphic respectively anti-holomorphic. 
However, one can 
define new fields $\mathcal J^a$ and $\overline{\mathcal J}{}^a$ 
\begin{eqnarray}
  \label{def_cur_04}
\begin{split}
  {\cal J}^a (z,\overline z) & = J^a(z)-
  {\textstyle \frac{1}{2}}\hspace{0.5pt}H^a{}_{bc} \, J^b(z) \, 
  {X}^c_R(\overline z)  \;, \\[1mm]
  \overline{\cal J}^a (z,\overline z) & = \overline J{}^a(\overline z)-
  {\textstyle \frac{1}{2}}\hspace{0.5pt}H^a{}_{bc} \, X_L^b( z) \,
  \overline J{}^c(\overline z)  \;
\end{split}
\end{eqnarray}
so that the three current correlators take the CFT form
\begin{eqnarray}
  \label{cur_cor_19}
\begin{split}
 \bigl\langle {\cal J}^a(z_1,\overline z_1)\, {\cal J}^b(z_2,\overline z_2)\, {\cal J}^c(z_3,\overline z_3) \bigr\rangle  
 &=  -i\:\frac{{\alpha'}^2}{8}\,H^{abc}\:
   \frac{1}{ z_{12}\, z_{23}\, z_{13}} \;, \\
  \bigl\langle \overline{\cal J}{}^a(z_1,\overline z_1)\, \overline{\cal J}{}^b(z_2,\overline z_2)\,
  \overline {\cal J}{}^c(z_3,\overline z_3) \bigr\rangle 
  &= +i\:\frac{{\alpha'}^2}{8}\,H^{abc}\:
   \frac{1}{ \overline z_{12}\, \overline z_{23}\, \overline z_{13}} \;.
\end{split}
\end{eqnarray}
The necessity of this redefinition  can already be understood from the 
two-dimensional equation of motion
$\partial\overline\partial X^a={1\over 2} H^a{}_{bc} \partial X^b
\overline\partial X^c$. Therefore, already at linear order the 
coordinate fields have to be adjusted to be consistent with a CFT description.
However, the deformation is still marginal, which means that  there are no
infinities leading to a renormalization group flow.

Writing the new currents as derivatives of corrected coordinates 
${\cal X}^a$, after three integrations the three-point function
of these coordinates can be computed as
\begin{eqnarray}
  \label{tpf_01aa}
  \bigl\langle {\cal X}^a(z_1,\overline z_1)\, {\cal X}^b(z_2,\overline z_2)\,   {\cal X}^{c}(z_3,\overline z_3) \bigr\rangle^{H}   
= \Theta^{abc} 
   \Bigl[ {\cal L} \Bigl( {\textstyle \frac{z_{12}}{z_{13}} } \Bigr)
   -  {\cal L} \Bigl( {\textstyle \frac{\overline z_{12}}{\overline z_{13}} } \Bigr) \Bigr]
\end{eqnarray}
with $\Theta^{abc}=\frac{{\alpha'}^2}{12}\: H^{abc}$ and 
\begin{equation}
{\cal L}(z)=
L(z) 
  + L \Bigl( 1-{1\over z} \Bigr) 
  + L \Bigl( {1\over 1-z} \Bigr)\quad {\rm where}\quad  L(z)={\rm Li}_2(z) +
  \frac{1}{ 2} \log (z) \log(1-z)\;   
\end{equation}
denotes  the (complex) Rogers dilogarithm\footnote{The theory of the 
complex Rogers dilogarithm is more involved than the real analogue. 
Some its features are collected in the appendix of \cite{Blumenhagen:2011ph}.}.
It satisfies the so-called fundamental  identity  $L(z)+L(1-z)=L(1)$.
The three-point function \eqref{tpf_01aa} should be considered
as the closed string genera\-lization of the second term in
\eqref{twopointboundary}. However, there are two essential differences:
\begin{itemize}
\item{For the closed string it is the three- and not the two-point function
which is corrected.}
\item{For the closed  string the Rogers dilogarithm  gives rise to
a non-trivial world-sheet dependence,
whereas for the open string only the essentially constant 
step-function appeared.}
\end{itemize} 
One can also compute the correction to the two-point function of two
coordinates. It reads 
\begin{equation}
   \delta_2 \bigl\langle X^a(z_1,\overline z_1) \, X^b(z_2,\overline z_2)  \bigr\rangle =
   \frac{{\alpha'}^2}{8} H^a{}_{pq}\, H^{bpq} \, \log |z_1-z_2|^2\;  \log\epsilon \;,
\end{equation}
where $\epsilon$ is a cut-off. 
Therefore, we explicitly see that the perturbation ${\cal S}_1$ ceases 
to be marginal at second order in the flux. The theory is no longer 
conformally invariant and starts to run according to the 
renormalization group flow equation for the inverse world-sheet metric $G^{ab}$,
which is  of the form 
\begin{equation}
  \mu\: \frac{\partial\hspace{0.5pt} G^{ab}}{ \partial \mu} =-\frac{\alpha'}{4} 
   H^a{}_{pq}\, H^{bpq}  \; .
\end{equation}
This  precisely agrees with equation \eqref{targeteom} for constant space-time metric,  $H$-flux and dilaton.

Even though in the framework of the Buscher rules, applying three
T-dualities on the $H$-flux background is questionable, 
on the level of the CFT, T-duality corresponds to a simple
asymmetric transformation of the world-sheet theory.
It is just a reflection of the right-moving coordinates. 
Since our  corrected fields ${\cal X}^a(z,\overline z)$  still admit
a split into a holomorphic and an anti-holomorphic piece,
we define T-duality on the  world-sheet action 
along the direction $\mathcal X^a$ as 
\begin{equation}
  \begin{array}{c}
  {\cal X}_L^a(z) \\[1mm]
  {\cal  X}_R^a(\overline z)   
  \end{array}
  \qquad
  \xrightarrow{\;\mbox{\scriptsize T-duality}\;}
  \qquad
  \begin{array}{c}
  +{\cal X}_L^a(z)\;, \\[1mm]
  -{\cal  X}_R^a(\overline z) \;.
  \end{array}
\end{equation}
Under a T-duality in all three directions,  momentum modes in the
$H$-flux background are mapped to winding modes in the $R$-flux
background. We are now interested in momentum modes in the
R-flux background  which are related via T-duality to
winding modes in the $H$-flux background. 
Therefore, the  three-point function in the $R$-flux background should
read
\begin{eqnarray}
  \label{tpf_01}
  \bigl\langle {\cal X}^a(z_1,\overline z_1)\, {\cal X}^b(z_2,\overline z_2)\,   {\cal X}^{c}(z_3,\overline z_3) \bigr\rangle^{R}   
= \Theta^{abc} 
   \Bigl[ {\cal L} \Bigl( {\textstyle \frac{z_{12}}{z_{13}} } \Bigr)
   +  {\cal L} \Bigl( {\textstyle \frac{\overline z_{12}}{\overline z_{13}} }
   \Bigr) \Bigr]\, ,
\end{eqnarray}
which just has a different relative sign between the holomorphic
and anti-holomorphic part. Here, we have the relation
$\Theta^{abc}=\frac{{\alpha'}^2}{12}\: R^{abc}$.

Up to linear order in the flux we can  write the energy-momentum tensor
as
\begin{equation}
  \label{def_emt_01}
  {\cal T}(z) = \frac{1}{\alpha'}\: \delta_{ab}:\! {\cal J}^a {\cal J}^b\!:\! (z) \;, \hspace{40pt}
  \overline{\cal T}(\overline z) = \frac{1}{\alpha'}\: \delta_{ab} :\! \overline{\cal J}{}^a\overline{\cal J}{}^b\!:\! (\overline z) \;.
\end{equation}
They give rise to  two copies of the Virasoro algebra with central charge
$c=3$ and ${\cal J}^a(\overline{\cal J}{}^a)$ is indeed a (anti-)chiral 
primary field with $h=1(\overline h=1)$.
Recall that the aim is to carry out a similar computation as for the open
string case, i.e. to evaluate string scattering amplitudes
for vertex operators and to see whether there is any sign
of a new space-time noncommutative/nonassociative product.
In the free theory the tachyon vertex operator
is a primary field  of conformal dimension $(h,\overline h)=(\frac{\alpha'}{4}\,
p^2,\frac{\alpha'}{4}\, p^2)$, and
in covariant quantization of the bosonic string
physical states are given by primary fields of conformal
dimension $(h,\overline h)=(1,1)$. 
The natural definition of the tachyon vertex operator 
for the perturbed theory is
\begin{equation}
  \label{def_vo7023}
  {\cal V}(z,\overline z) 
= \, :\!\exp \bigl( i\hspace{0.5pt} p \cdot ({\cal X}_L +
  {\cal X}_R) \bigr) \!: \;.
\end{equation}
One can compute
\begin{eqnarray}
\begin{split}
   {\cal T}(z_1)\, {\cal V}(z_2,\overline z_2) 
   &=
   \frac{1}{(z_1-z_2)^2} \, \frac{\alpha' p\cdot p}{4} \,   {\cal V}(z_2,\overline z_2)  +\frac{1}{z_1-z_2} \, \partial {\cal V}(z_2,\overline z_2) 
   +{\rm reg.} \;,
\end{split}
\end{eqnarray}
and analogously for the anti-holomorphic part.
This means that the vertex operator \eqref{def_vo7023} is  primary and 
has conformal dimension $(h,\overline h)=(\frac{\alpha'}{4}\,
p^2,\frac{\alpha'}{4}\, p^2)=(1,1)$. It is therefore a physical quantum
state of the  deformed theory.
For the correlator of three tachyon vertex operators one obtains
\begin{equation}
  \label{threetachyonb}
   \bigl\langle \,{\cal V}_1 \,{\cal V}_2 \,{\cal V}_3 \,\bigr\rangle^{H/R}
   =\frac{\delta(p_1+p_2+p_3)}{\vert z_{12}\,z_{13}\,z_{23}\vert^2}\,  \exp\Bigl[ -i\hspace{0.5pt}\Theta^{abc}\, p_{1,a} p_{2,b} p_{3,c}  \bigl[
  {\cal L}   \bigl( {\textstyle \frac{z_{12}}{z_{13}} }\bigr) \mp {\cal L}
   \bigl({\textstyle \frac{\overline z_{12}}{ \overline z_{13}}}\bigr)\bigr] \Bigr]_{\Theta} \,,
\end{equation}
where $[\ldots]_{\Theta}$ indicates that the result is valid only up to linear order in $\theta$.
The full string scattering amplitude of the integrated tachyon vertex
operators then becomes
\begin{eqnarray}
  \label{threetacyon}
\begin{split}
  \bigl\langle\, \mathcal T_1\: \mathcal T_2\:\mathcal  T_3\, \bigr\rangle^{H/R}
  &= \int \prod_{i=1}^3  d^2 z_i\, \delta^{(2)}(z_i-z_i^0) \, \delta(p_1+p_2+p_3) \times \\
 & \hspace{40pt}
 \exp\Bigl[ -i\hspace{0.5pt}\Theta^{abc}\, p_{1,a} p_{2,b} p_{3,c}  \bigl[
  {\cal L}   \bigl( {\textstyle \frac{z_{12}}{z_{13}} }\bigr) \mp {\cal L}
   \bigl({\textstyle \frac{\overline z_{12}}{ \overline z_{13}}}\bigr)\bigr] \Bigr]_{\Theta}.
\end{split}
\end{eqnarray}

Let us now study the behavior of \eqref{threetachyonb}  under
permutations of the vertex operators. 
Before applying momentum conservation, 
the three-tachyon amplitude for a permutation $\sigma\in S_3$ of
the vertex operators can be computed 
using the  relation $L(z)+L(1-z)=L(1)$. 
With $\epsilon=-1$ for the $H$-flux and $\epsilon=+1$ for the $R$-flux, 
one finds
\begin{equation}
  \label{phasethreeperm}
\bigl\langle \, {\cal V}_{\sigma(1)}   {\cal V}_{\sigma(2)}  {\cal V}_{\sigma(3)}  \bigr\rangle^{H/R}= \exp\Bigl[ \,i\left({\textstyle \frac{1+\epsilon}{ 2}}\right)  \eta_\sigma\,  \pi^2\,  \Theta^{abc}\, p_{1,a} 
  \,p_{2,b} \,p_{3,c} \Bigr]
  \bigl\langle {\cal V}_1\,  {\cal V}_2\,  {\cal V}_3  \bigr\rangle^{H/R} \;,
\end{equation}
where in addition $\eta_{\sigma}=1$ for an odd permutation and 
$\eta_{\sigma}=0$ for an even one. 
One observes that for  $H$-flux the phase is always trivial while 
for  $R$-flux  a non-trivial phase may appear.
Recall that our analysis is only reliable up to linear order in $\Theta^{abc}$.

Note that it is  non-trivial  that this phase is independent
of the world-sheet coordinates, which can be traced back  to
the form of the fundamental identity of $L(z)$. 
For this reason, it can be thought of as a property of the underlying target 
space.
Indeed, the phase in \eqref{phasethreeperm} can be recovered from 
a new three-product on the space of  functions
$V_{p_n}(x)=\exp( i\, p_n \cdot x )$  which is defined as 
\begin{equation}
  \label{threebracketexp}
   V_{p_1}(x)\,\tri\, V_{p_2}(x)\, \tri\, V_{p_3}(x)\stackrel{\rm def}{=} 
  \exp\Bigl( -i \,{\textstyle \frac{\pi^2}{2}}\, \Theta^{abc}\,
   p_{1,a}\, p_{2,b}\, p_{3,c} \Bigr) V_{p_1+p_2+p_3}(x)\; .
\end{equation}
However, in CFT correlation functions operators
are understood to be radially ordered and so changing the order
of operators should not change the form of the amplitude.
This is known as crossing symmetry and is also called 
the associativity of the operator product expansion.
In the case of the $R$-flux background, this
is reconciled by applying momentum conservation leading to
\begin{equation}
  p_{1,a} \,p_{2,b}\, p_{3,c}\,\Theta^{abc} = 0
  \hspace{40pt}{\rm for}\hspace{40pt} p_3=-p_1-p_2 \;.
\end{equation}
Therefore, scattering amplitudes of three tachyons do not receive
any corrections at linear order in $\Theta$  both for the $H$- and
 $R$-flux.  

The tri-product \eqref{threebracketexp} can be generalized to 
 more generic functions as
\begin{equation}
\label{threebracketcon}
   f_1(x)\,\tri\, f_2(x)\, \tri\, f_3(x) \stackrel{\rm def}{=} \exp\Bigl(
   {\textstyle {\pi^2\over 2}}\, \Theta^{abc}\,
      \partial^{x_1}_{a}\,\partial^{x_2}_{b}\,\partial^{x_3}_{c} \Bigr)\, f_1(x_1)\, f_2(x_2)\,
   f_3(x_3)\Bigr|_{x} \;,
\end{equation}
where we used the notation $(\ )\vert_x=(\ )\vert_{x_1=x_2=x_3=x}$.
This is to be compared with the $\star$-product \eqref{starproduct}
and can be thought of as a possible closed string generalization
of the open string noncommutative structure.
For the tri-bracket of  the coordinates $x^a$ 
one gets $\bigl [x^a,x^b,x^c \bigr] = 3\pi^2\, \Theta^{abc}$.

This result generalizes to the $N$-tachyon amplitude, where
the  relative phase can be described by the following deformed product
\begin{equation}
\label{triprodn}
  f_1(x)\, \tri_N\,  \ldots \tri_N\,  f_N(x) \stackrel{\rm def}{=} 
   \exp\left[ {\textstyle {\pi^2\over 2}} \Theta^{abc}\!\!\!\!\! \sum_{1\le i< j < k\le N}
     \!\!\!\!  \, 
      \partial^{x_i}_{a}\,\partial^{x_j}_{b} \partial^{x_k}_{c} \right]\, 
   f_1(x_1)\, f_2(x_2)\ldots
   f_N(x_N)\Bigr|_{x} \;, 
\end{equation}
which has the peculiar feature that  the phase becomes trivial 
after taking momentum conservation into account or equivalently 
\begin{equation}
  \int d^n x \,f_1(x)\, \tri_N\,  f_2(x)\, \tri_N \ldots \tri_N\,  f_N(x) 
   = \int d^n x\, f_1(x)\,f_2(x)\, \ldots \,  f_N(x)\, .\quad 
\end{equation}
We conclude that for constant closed string $R$-flux, 
a non-vanishing tri-product 
\eqref{threebracketcon} and \eqref{triprodn} is
consistent with the main axioms of a conformal field theory.
Below the integral it only gives boundary terms and therefore
does not change the equations of motion. Therefore, we have identified
a non-trivial nonassociative deformation of the underlying target space,
which is not visible in CFT. Whether this deformation is ``real'' can only
be decided, if one goes beyond conformal field theory, i.e.
the usual perturbative approach to string theory.
Some simple ideas in this direction will be discussed
in the following final part of this lecture.

\vspace{0.5cm}
\noindent
{\bf Beyond constant $\Theta^{abc}$ flux}
\vspace{0.2cm}

When the lecture was given, the question arose what 
happens for non-constant $\Theta$-flux and whether it helps
to decide about the relevance of this entire approach.
Thus, the question is how the open string analysis from
the end of section \ref{sec_open} generalizes to the closed string.
Since nongeometric fluxes are appropriately described in double
field theory (DFT), the analysis should be performed
in this framework. This was indeed done in \cite{Blumenhagen:2013zpa}. 
For self-consistency
of this lecture, here we will not work in the most general
DFT setting but instead   present some of the 
essential results in a simplified form.

For the open string, it happened that both cyclicity and associativity
hold up to boundary terms precisely if the string equations
of motion were satisfied for the background.
Let us consider  the tri-product \eqref{threebracketcon} of three functions
below an integral on  a manifold with curved metric $g_{ab}$
\eq{
    \int dx \,\sqrt{-g}\,e^{-2\phi}\, f_1 \triangle f_2 \triangle f_3 =
    \int dx \,\sqrt{-g}\, e^{-2\phi}\, f_1 \,f_2 \, f_3 +
     {\pi^2\over 2} \int dx\, \sqrt{-g} \,e^{-2\phi}\, \Theta^{abc}\, 
 \partial_{a} f_1 \,\partial_{b} f_2 \partial_{c} f_3 + \ldots\, .
}
Performing an integration by parts for the second term, we get
\eq{
      \int dx\, \sqrt{-g} e^{-2\phi}\,\Theta^{abc}\, 
 \partial_{a} f_1 \,\partial_{b} f_2 \partial_{c} f_3 =
  \int dx\, \partial_a \Big( \sqrt{-g}&\, e^{-2\phi} \,\Theta^{abc}\, 
 f_1 \,\partial_{b} f_2 \partial_{c} f_3 \Big)\\
  &- \int dx\, \partial_a \Big( \sqrt{-g} \, e^{-2\phi}\, \Theta^{abc} \Big)\, 
 f_1 \,\partial_{b} f_2 \partial_{c} f_3 
}
which gives a boundary term for 
\eq{
         \partial_a \Big( \sqrt{-g} \, e^{-2\phi} \,\Theta^{abc} \Big)=0\, .
}
This is precisely the leading order equation of motion for
the $H$-flux, i.e. for $\Theta^{abc}=g^{aa'}g^{bb'}g^{cc'} H_{a'b'c'}$.
Therefore, for such a tri-product, its effect at linear order 
(nonassociativity described by a tri-bracket) gives a boundary
term. In \cite{Blumenhagen:2013zpa} it was shown 
that this behavior also holds for a doubled flux $F^{ABC}$ 
which includes  also the gravitational part. 
         
However, this is actually {\it not} the flux, for which 
the tri-product appeared in the CFT. There, we argued that
it is the $R$-flux, i.e. $\Theta^{abc}=R^{abc}$, which by itself
is already an anti-symmetric three-vector so that no metric factors
are involved. Therefore, in this sense such a tri-product is
rather of topological type and should involve Bianchi identities
instead of equations of motion. In \cite{Blumenhagen:2013zpa}
this was shown to be essentially correct in the DFT framework.
However, in DFT the $O(D,D)$ symmetry requires that with
$R^{abc}=3\, \partial^{[a} \beta^{bc]}$ the complete tri-product
takes the form
\eq{
   f_1 \triangle f_2 \triangle f_3 =
   f_1 \,f_2 \, f_3 +
     {\pi^2\over 2} \Big(R^{abc}\, 
 \partial_{a} f_1 \,\partial_{b} f_2 \partial_{c} f_3 + Q_a{}^{bc}\, 
 \big( \tilde\partial^{a} f_1 \,\partial_{b} f_2 \partial_{c} f_3
  +{\rm cycl}_{f_1,f_2,f_3}\big)\Big)\, .
}
Employing the strong constraint \eqref{strong_c} between $\beta$ and
the $f_i$, the entire second term on the right hand side 
vanishes and the tri-product becomes trivial. Thus, in DFT
one can only  get a non-trivial tri-product if the strong constraint
is weakened. A more elaborate discussion can be 
found in \cite{Blumenhagen:2013zpa}. 

This reflects the status of the emergence of nonassociativity
in closed string theory and it is fair to say that this issue
is still not completely settled and more work is needed to put
together the bits and pieces that were revealed so far.

\begin{acknowledgement}
I would like to thank my collaborators Andre Betz,  Andreas Deser, Michael Fuchs, Xin Gao, Falk
Hassler, Daniela Herschmann, Dieter L\"ust, Erik Plauschinn, Felix Rennecke, 
Christian Schmid, Pramod Shukla and Rui Sun 
for the many discussions
on the subject presented in this mini course.
Moreover, I thank the organizers of the Corfu Workshop on
Noncommutative Field Theory and Gravity
September 8 - 15, 2013, for creating
such a stimulating  atmosphere and the participants for asking so many
good questions.
\end{acknowledgement}

\end{document}